\documentclass[%
 reprint,
 amsmath,amssymb,
 aps,
]{revtex4-2}
\usepackage{setspace}
\usepackage{graphicx}
\usepackage{dcolumn}
\usepackage[colorlinks=true,linkcolor=blue,citecolor=blue,urlcolor=blue]{hyperref}
\usepackage{bm}
\usepackage[mathlines]{lineno}

\usepackage{soul}
\usepackage{yfonts}
\usepackage{amsmath}
\usepackage{xcolor}
\usepackage{float}
\usepackage{amsfonts}
\usepackage{amsthm}

\begin{document}

\preprint{APS/123-QED}

\title{Enhancement in phase sensitivity of SU(1,1) interferometer with Kerr state seeding}
\author{Priyanka Sharma}%
\author{Aviral K. Pandey}%
\author{Gaurav Shukla}%
\author{Devendra Kumar Mishra}%
\email{kndmishra@gmail.com}
\affiliation{%
 Department of Physics, Institute of Science, Banaras Hindu University, Varanasi-221005, India}%

\vspace{10pt}

\date{\today}

\begin{abstract}
A coherent seeded SU(1,1) interferometer provides a prominent technique in the field of precision measurement. We theoretically study the phase sensitivity of SU(1,1) interferometer with Kerr state seeding under single intensity and homodyne detection schemes. To find the lower bound in this case we calculate the quantum Cramér-Rao bound using the quantum Fisher information technique. We found that, under some conditions, the Kerr seeding performs better in phase sensitivity compared to the well-known vacuum and coherent seeded case. We expect that the Kerr state might act as an alternative non-classical state in the field of quantum information and sensing technologies.
\end{abstract}

\maketitle


\section{Introduction}\label{section1}

Interferometry, a technique based on wave interference, played a crucial part in science and technology. It has been an indispensable part of precision measurement ever since its inception. For precision measurement of parameters \cite{1976Helstrom}, which are not measurable directly via conventional techniques, phase estimation by optical interferometers plays an important role \cite{doi:10.1080/00107510802091298, RevModPhys.90.035005,2015PrOpt..60..345D, Lawrie2019, 1976Helstrom, sharma2023superresolution}. Due to vacuum fluctuations, optical interferometers possess a lower bound in the precision measurement called shot noise limit (SNL) and equal to $1/\sqrt{n}$ where $n$ is the mean photon number inside the interferometer. One can beat the SNL using nonlinear light-matter interactions like, nonclassical states \cite{Shukla:21, Shukla:22, Mishra_2021, SHUKLA2024100200, shukla2023improvement} and nonlinear materials inside the interferometer \cite{Chekhova:16}. Beating SNL one can reach the Heisenberg limit (HL), i.e., $1/n$ \cite{PhysRevA.94.063840}.

Based on the transformations of the optical fields, optical interferometers are divided into SU(2) and SU(1,1) classes \cite{PhysRevA.33.4033}. SU(2) interferometers, like Mach-Zehnder interferometer (MZI) and Michelson interferometer, are based on the passive type of beam splitters and follow the SU(2) transformation in linear wave mixing. In 1986, Yurke \textit{et al.} \cite{PhysRevA.33.4033} proposed the SU(1,1) interferometer in which they replaced the beam splitter with the optical parametric amplifiers (OPAs) and showed that with vacuum, SU(1,1) interferometer can achieve the HL in the lossless case. The name SU(1,1) stems from the Bogoliubov transformations, related to the SU(1,1) group, of the optical fields involved in parametric processes. Plick \textit{et al.} \cite{Plick1_2010} showed that single seeding of the coherent state improves the sensitivity of the SU (1,1) interferometers and surpasses the SNL. Subsequently, Li \textit{et al.} showed by using parity detection SU(1,1) interferometer can achieve the HL \cite{PhysRevA.94.063840}.  Marino \textit{et al.} \cite{PhysRevA.86.023844} showed the effect of the optical losses on the sensitivity performance. Interestingly, Hudelist \textit{et al.} pointed out that the signal-to-noise ratio (SNR) of the SU(1,1) interferometer is about 4.1 dB higher than that of the MZI under the same phase sensing intensity \cite{hudelist2014quantum}. This implies that the role of a nonlinear process are very much important in improving the precision of the measurement.

Recent studies showed that the implementation of the Kerr medium inside the SU(1,1) interferometer improves the performance of the precision measurement \cite{PhysRevA.105.033704}. For SU(2) interferometer, Yadav \textit{et al.} find the improvement in the phase sensitivity by applying the squeezed Kerr state as the input in lossless as well as in lossy conditions \cite{yadav2023quantumenhanced}. The reason behind the improvement in the results are inherent properties of the Kerr state. The Kerr effect generates quadrature squeezing but does not modify the input field photon statistics, i.e., it remains Poissonian \cite{PhysRevA.34.3974}. It possesses the squeezing in an oblique direction neither in the direction of phase quadrature nor the direction of amplitude quadrature \cite{sizmann1999v}. 

Based on these results we are motivated and propose the Kerr seeding instead of a coherent state in the degenerate SU(1,1) interferometer. Here, we calculate the phase sensitivity using single intensity (SI) and homodyne detection (HD) schemes and compare the results with the coherent state-seeded case. In order to obtain the lower bound on the precision measurement we calculate the Cramér-Rao bound (QCRB) using the quantum Fisher information (QFI) technique \cite{PhysRevA.99.042122, zeng2023effect}.

The paper is organized as follows. In section \ref{section2}, we discuss the generation of the Kerr state, the basics of SU(1,1) interferometer and QCRB with the method of QFI. Section \ref{section3} contains observation with Kerr state seeding with two detection schemes in lossless cases. Section \ref{section4} is the observation for the lossy case. Finally, In Section \ref{section5}, we conclude our results.

\section{Generation of Kerr state and  basics of SU(1,1) interferometer}\label{section2}
 
\subsection{Generation of Kerr state}

The Kerr effect is also known as a quadratic electro-optic effect. This is a change in the refractive index of a material medium in response to an applied electromagnetic field. In the Kerr medium, the electromagnetic field interacts with the material medium having third-order non-linearity. Here the refractive index is intensity-dependent \cite{Agarwal_2012, gerry_knight_2004}. Therefore, Hamiltonian $\hat{H}$, of this quantum mechanical system, can be written as \cite{Agarwal_2012, gerry_knight_2004} 
\begin{equation} \hat{H}=\hslash{\omega}\hat{a}^\dagger\hat{a}+\hslash{\chi^{(3)}}\hat{a}^{\dagger2}\hat{a}^{2}, 
\end{equation}
where $\omega$ is the frequency, $\hslash$ is the plank constant \cite{BUNKER2019106594},  $\hat{a}~(\hat{a}^\dagger)$ is the annihilation (creation) operator of the oscillator and $\chi^{(3)}$ is the third-order susceptibility of the Kerr medium. For the generation of the Kerr state, we inject the coherent state into the Kerr medium. Therefore, injecting the light beam in a  coherent state $|\alpha\rangle$ through the material having $\chi^{(3)}$ non-linearity results Kerr state which can be written as
 \begin{equation}
     |\psi_{K}\rangle=\hat{U}(\gamma)|\alpha\rangle. \label{eqkerr}
 \end{equation}
Where $\hat{U}(\gamma)$ is the operator associated with the Kerr medium, defined as
\begin{equation}
    \hat{U}(\gamma)=\exp[-i\gamma\hat{n}(\hat{n}-1)],
\end{equation}
 where $\hat{n}$ $(=\hat{a}^{\dagger}\hat{a})$ is the photon number operator and,
 \begin{equation}
     \gamma=\chi^{(3)}{l}/u,\label{K19}
 \end{equation}
 with $l$ being the length of the Kerr medium and $u$ being the velocity of the electromagnetic field into the Kerr medium. As we can see from Eq.\eqref{K19}, $\gamma$ tells about the interaction time of the electromagnetic field with $\chi^{(3)}$ material medium. Hence, we can call $\gamma$ as the Kerr interaction coefficient.

Let, $\hat{a}$ be the field operator for the coherent state, then the field operator associated with Kerr state can be written as
\begin{eqnarray}
\hat{a}(\gamma)\equiv\hat{U}^{\dagger}(\gamma)\hat{a}\hat{U}(\gamma)=e^{-2i\gamma\hat{a}^{\dagger}\hat{a}}\hat{a}\label{28l}.
\end{eqnarray}
We use this operator in our calculations in order to find the general result for the phase sensitivity using the Kerr state as one of the inputs of the SU(1,1) interferometer.

\subsection{Basics of SU(1,1) Interferometer}
\begin{figure}
\includegraphics[width=8.2cm, height=5.2cm]{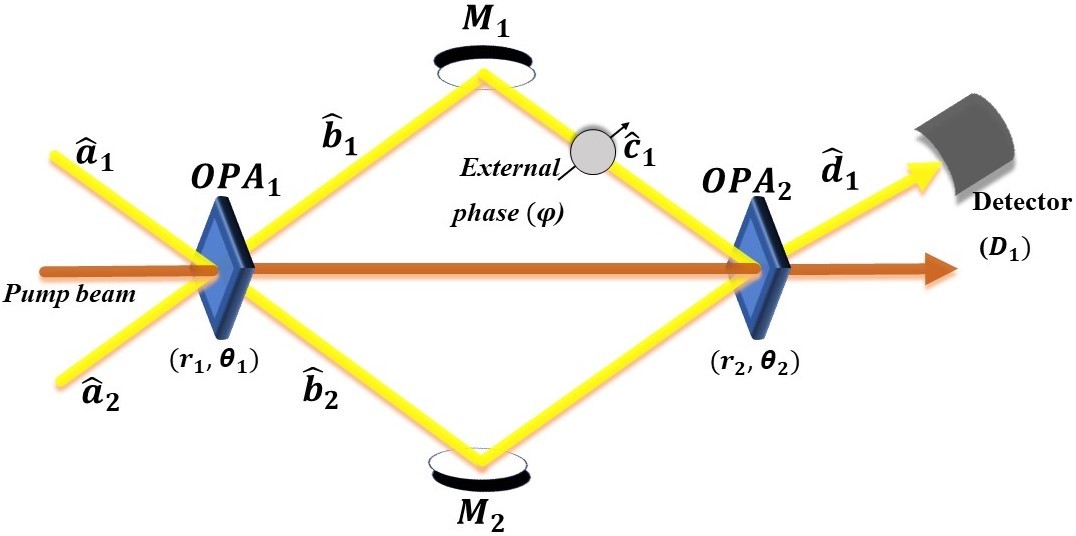}
\caption{\label{fig_1}The schematic diagram of SU(1,1) interferometer in lossless case. It consists of two optical parametric amplifiers, OPA-1 and OPA-2 ($r_1,r_2$ and $\theta_1, \theta_2$ are the squeezing parameter and phase of squeezing respectively), two mirrors($M_1, M_2$), detector($D_1$) and $\phi$ is the external phase introduced in the path of the beams. The pump beam between the two OPAs is used as a reference beam.}
 \end{figure}

The setup of the SU(1,1) interferometer is shown in Fig. \ref{fig_1}. It consists of two degenerate optical parametric amplifiers, $OPA_1$ ($r_1,\theta_1$) and $OPA_2$ ($r_2,\theta_2$), where $r$ and $\theta$ are the amplitude and phase of the squeezing, respectively. With the help of the detector $D_1$ we measure the phase change, $\phi$, in the arm of the interferometer, as shown in Fig. \ref{fig_1}.

The operator transformation in the SU(1,1) interferometer follows the SU(1,1) algebra. From the Bogoliubov transformation, the input and output of the OPA can be written as 
\begin{equation}
\begin{split}
\hat{a}_{out_1}=\cosh{r_1}\hat{a}_{in_1} + e^{i\theta} \sinh{r_1}\hat{a}_{in_2}^{\dagger},\\
\hat{a}_{out_2}=\cosh{r_1}\hat{a}_{in_2} + e^{i\theta} \sinh{r_1}\hat{a}_{in_1}^{\dagger}.
\end{split}\label{b12}
\end{equation}
Where $\theta$ is the phase of the OPA. The phase shift $\phi$ in the upper arm is modelled by the unitary operator $\hat{U}_\phi = e^{i\hat{n}\phi}$ where $\hat{n}(= \hat{a}^\dagger \hat{a})$ is the number operator.

Now, in order to perform the measurement on the output port of the interferometer, we must have an observable corresponding to the detection scheme. Let us consider $\hat{A}(\phi)$ as an operator that contains the information of the unknown phase $\phi$ to be estimated. From the standard error propagation formula the phase sensitivity $\Delta\phi$ is written as
\begin{equation}
\Delta\phi = \frac{\Delta\hat{A}(\phi)}{\left|\frac{\partial\langle\hat{A}(\phi)\rangle}{\partial\phi}\right|}\label{8a},
\end{equation}
where $\Delta\hat{A}(\phi) = \sqrt{\langle\hat{A}^2\rangle -\langle\hat{A}\rangle^2}$ is the standard deviation of $\hat{A}(\phi)$.

In this paper, we will use the single intensity (SI) and homodyne detection (HD) schemes. The corresponding operators are defined as
\begin{equation}
\hat{A}_{si}(\phi) = (\hat{d}_1^\dagger \hat{d}_1),\label{8b}
\end{equation} 
\begin{equation}
\hat{A}_{hd}(\phi)=\frac{(\hat{d}_1+\hat{d}_1^\dagger)}{\sqrt{2}},\label{8c}
\end{equation} 
respectively. Where $\hat{d_1}(\hat{d_1}^{\dagger})$ is the annihilation (creation) operator on the output port as shown in Fig. \ref{fig_1}.

\subsection{Quantum Fisher Information(QFI) and Quantum Cramér-Rao Bound(QCRB)} 

The QCRB gives the lower bound in the phase estimation process. The QCRB for the multi-parameter, say $\phi_1$ and $\phi_2$, is estimated by using the QFI matrix \cite{zeng2023effect}. For the estimation of the sum phase $\phi_{s} (= \phi_1 + \phi_2)$ and difference phase $\phi_{d} (= \phi_1 - \phi_2)$ the QFI matrix is defined as
\begin{equation}
   \mathcal{F}_Q =\begin{pmatrix}
    \mathcal{F}_{ss}&\mathcal{F}_{sd}\\
    \mathcal{F}_{ds}&\mathcal{F}_{dd}.
\end{pmatrix}
\end{equation}
The element of the matrix is defined as 
\begin{equation}
    F_{ij}=4 (\langle \hat{g}_i\hat{g}_j \rangle - \langle \hat{g}_i \rangle \langle \hat{g}_j \rangle),
\end{equation}
where $i, j = s, d$ and $\hat{g}_{i,j} = ( \hat{n}_1 \pm \hat{n}_2 ) / 2$.
For the SU(1,1) interferometer, we are usually interested in the estimation of only sum phase $\phi_s$ and here in our case $\phi_s$ is $\phi$. Therefore, for estimation of only $\phi_s$, QFI is given by
\begin{equation}
 \mathcal{F}_Q = F_{ss}-\frac{F_{sd}F_{ds}}{F_{dd}}.
\end{equation}
Can be written as 
\begin{equation}
    \mathcal{F}_Q = 4\frac{\langle\Delta^2\hat{n}_1\rangle \langle\Delta^2\hat{n}_2\rangle-Cov[n_1,n_2]^2}{\langle\Delta^2\hat{n}_1\rangle+\langle\Delta^2\hat{n}_2\rangle-2 Cov[\hat{n}_1,\hat{n}_2]} \label{F}
\end{equation}
Where,
$\langle\Delta^2\hat{n}_k\rangle=\langle\hat{n}^2_k\rangle-\langle\hat{n}_k\rangle^2$ and $Cov[n_1,n_2]=\langle \hat{n}_1\hat{n}_2\rangle-\langle \hat{n}_1\rangle\langle\hat{n}_2\rangle$.

Thus, QCRB is \cite{PhysRevLett.111.173601}
\begin{equation}
\Delta\phi_{Q} = \frac{1}{\sqrt{{\mathcal{F}_Q}}}\label{qcrb}.
\end{equation}
This gives us the ultimate precision achievable on the estimation of $\phi$ independent of a quantum measurement.

\section{Observation with Kerr state seeding as input in lossless case}\label{section3}

In this section, we calculate the phase sensitivity in lossless case using the Kerr seeding (Eq. \eqref{eqkerr}). In this case, output operator can be written as
\begin{equation}
    \begin{split}
        \hat{d_1} = \hat{a}_1 e^{-i (\theta_1 + 2\gamma a_1^\dagger a_1)} \left(e^{i (\theta_1 + \phi)} \cosh(r_1) \cosh(r_2) \right.\\
        \left. + e^{i\theta_2} \sinh(r_1) \sinh(r_2)\right) + \left(e^{i (\theta_1 + \phi)} \cosh(r_2) \sinh(r_1) \right. \\
        \left. + e^{i\theta_2} \cosh(r_1) \sinh(r_2)\right) \hat{a}_2^\dagger.
    \end{split}
\end{equation}
Where $\hat{a}_1$ and $\hat{a}_2$ are annihilation operators for the coherent state and vacuum state, respectively. In order to calculate the phase sensitivity we use Eq. (\ref{8a}). For the case of SI detection, we use the operator defined in Eq. (\ref{8b}) and it gives us 
\begin{equation}
\Delta\phi_{si} = \frac{\sqrt{(\langle\hat{d}_1^{\dagger2}\hat{d}_1^{2}\rangle+\langle\hat{d}_1^{\dagger}\hat{d}_1\rangle) - \langle\hat{d}_1^\dagger\hat{d}_1 \rangle^2}}{\left|\frac{\partial\langle\hat{d}_1^\dagger\hat{d}_1 \rangle}{\partial\phi}\right|},
\end{equation}
where the expectation values of $\langle\hat{d}_1^{\dagger2}\hat{d}_1^{2}\rangle$ and $\langle\hat{d}_1^{\dagger}\hat{d}_1\rangle$ are defined in Appendix \ref{appendix A}.

Similarly for the HD case, using Eq. (\ref{8c}), phase sensitivity can be written as 
\begin{equation}
\Delta\phi_{hd} = \frac{\sqrt{{(\langle\hat{d}_1^{2}\rangle+\langle\hat{d}_1^{\dagger2}\rangle+2\langle\hat{d}_1^{\dagger}\hat{d}_1\rangle+1)} -{(\langle\hat{d}_1\rangle+\langle\hat{d}_1^\dagger\rangle)}^2}}{\left|\frac{\partial{(\langle\hat{d}_1\rangle+\langle\hat{d}_1^\dagger\rangle)}}{\partial\phi}\right|}.
\end{equation}
Where expectation values of first moment and second moment of $\hat{d}_1$, i.e., $\langle\hat{d}_1\rangle$ and $\langle\hat{d}_1^{2}\rangle$ is given in Appendix \ref{appendix A}.

\begin{figure}[H]
\includegraphics[width=8.7cm, height=6.2cm]{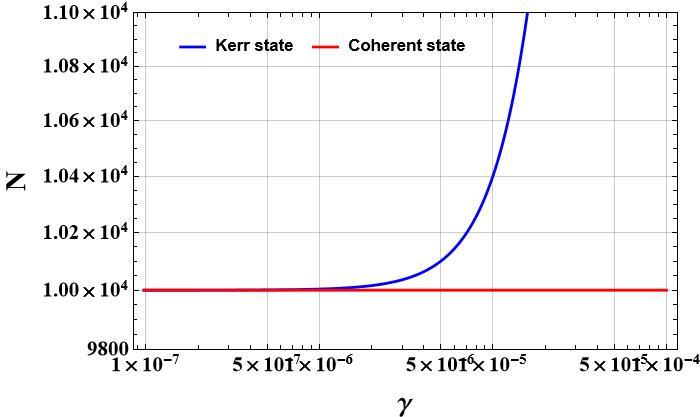}
\caption{\label{fig_2}The figure shows the variation of input photon number $N$ versus $\gamma$ for the value of $\alpha=100$.}
\end{figure}

Since the interaction of the light with $\chi^{(3)}$ type of material is very small it implies that $\gamma<<1$. This gives us the freedom to approximate the operator as 
\begin{equation}
e^{-2i\gamma \hat{a}^{\dagger}\hat{a}}=(1-2i\gamma\hat{a}^{\dagger}\hat{a}),\label{a1}
\end{equation}
and we use this approximation in our calculations. It is well known that the mean photon number in Kerr states and coherent states are equal \cite{PhysRevA.49.2033}. However, here, we take the approximation therefore, we should upper bound the value of $\gamma$ and for this purpose we plot mean photon number versus $\gamma$ for Kerr state ($N_{Kerr} = \alpha^2+4 \alpha^4 \left(1+\alpha^2\right) \gamma^2$) and coherent state ($N_{CS} = \alpha^2$) as shown in Fig. (\ref{fig_2}). In Fig. (\ref{fig_2}) for $\alpha=100$, the mean photon number of coherent state and Kerr state are equal up to $\gamma\approx1\times10^{-6}$. Therefore in our further discussion we upper bound the value of $\gamma$ with $\approx1\times10^{-6}$. This ensures that we have equal mean photons in both the Kerr state and coherent state.

To find the lower bound of the phase sensitivity we calculate QCRB for the Kerr state and coherent state (given in Eq. (\ref{qcrb})). For our present case, QCRB of the system is written as
\begin{equation}
    \Delta\phi_Q = \frac{1}{2}\sqrt{\frac{\langle\Delta^2(\hat{c}_1^\dagger\hat{c}_1)\rangle+\langle\Delta^2(\hat{b}_2^\dagger\hat{b}_2)\rangle-2 Cov[(\hat{c}_1^\dagger\hat{c}_1),(\hat{b}_2^\dagger\hat{b}_2)]}{\langle\Delta^2(\hat{c}_1^\dagger\hat{c}_1)\rangle \langle\Delta^2(\hat{b}_2^\dagger\hat{b}_2)\rangle-Cov[(\hat{c}_1^\dagger\hat{c}_1),(\hat{b}_2^\dagger\hat{b}_2)]^2}}, \label{F2}
\end{equation}
where required expectation values are given in Appendix \ref{appendix B}.

\begin{figure}
\includegraphics[width=8.5cm, height=6.5cm]{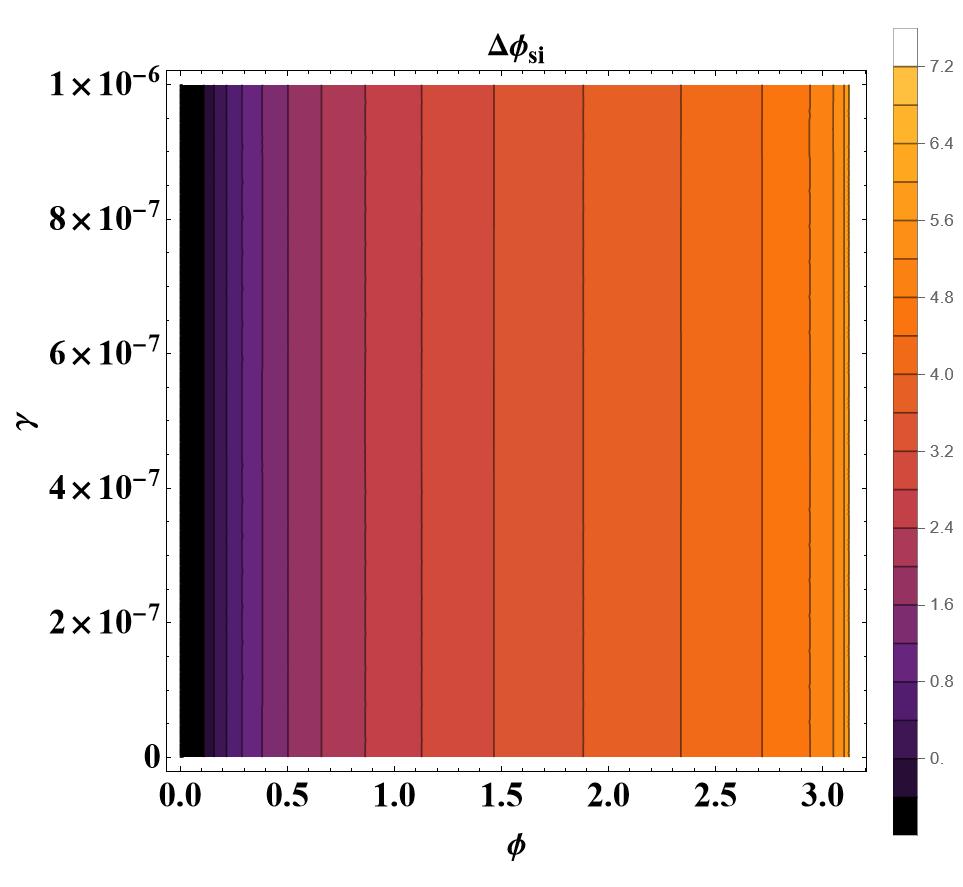}
\caption{\label{fig_3a}The contour plot shows the variation of $\Delta\phi_{si}$, $\gamma$ and $\phi$. From here we can see that in the case of SI scheme $\Delta\phi_{si}$ is independent from $\gamma$. Other parameters are $r_1 = r_2 = 2.0,~\theta_1=0,~\theta_2 = \pi,$ and $\alpha=100$.}
\end{figure} 

\begin{figure}[H]
\includegraphics[width=8.5cm, height=6.5cm]{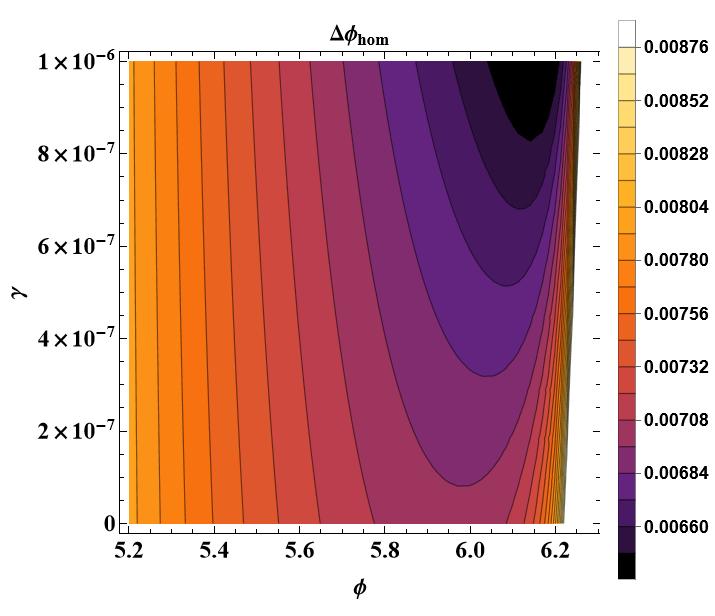}
\caption{\label{fig_3}The contour plot shows the variation of $\Delta\phi$, $\gamma$ and $\phi$ for homodyne detection scheme. From here we are getting optimum range of $\phi$ from $5.9$ to $6.19$ for the values of $r_1 = r_2 = 2.0,~\theta_1=0,~\theta_2=\pi,$ and $\alpha = 100$.}
\end{figure} 

To understand the effect of $\gamma$ on phase sensitivity we see the variation of $\Delta \phi$ with $\gamma$ and $\phi$ for SI and HD scheme as shown in Fig. \ref{fig_3a} and Fig. \ref{fig_3}, respectively. We have considered $\alpha=100$, $\gamma=1\times10^{-6}$. For simplicity, we consider the conventional SU(1,1) interferometer setup situation in which we choose equal amplitude of squeezing/anti-squeezing ($r_1 = r_2 = 2$) with $\pi$ phase difference between the two OPAs ($\theta_1=0$ and $\theta_2=\pi$). From Fig.  \ref{fig_3a} we can see that phase sensitivity is independent of the $\gamma$ in the SI scheme. While, Fig. \ref{fig_3}, shows the variation in phase sensitivity with $\gamma$ in the HD scheme. Therefore, further, we discuss the dependencies of phase sensitivity on $\gamma$ in different scenarios only in the HD scheme.  

\begin{figure}[H]
\includegraphics[width=8.7cm, height=6cm]{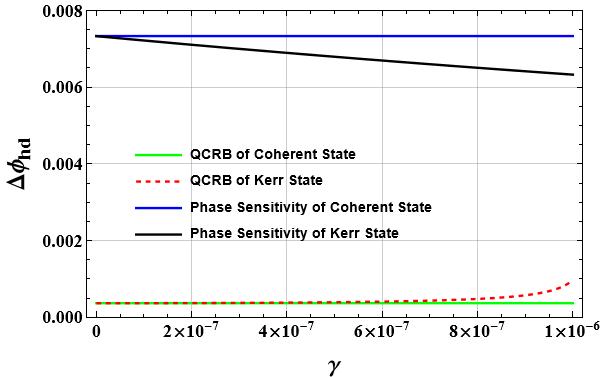}
\caption{\label{fig_5}The figure shows the variation of $\Delta\phi$ versus $\gamma$ for homodyne detection with the values $r_1 = r_2 = 2.0,~\phi=6.15,~\theta_1=0, ~\theta_2=\pi$, and $\alpha=100$.}
\end{figure}

In order to compare the phase sensitivity with coherent state seeding and Kerr state seeding in the HD scheme we plot Fig. \ref{fig_5}. From Fig. \ref{fig_5} we can see the improvement in phase sensitivity in the case of Kerr seeding. We can also see that the Kerr state seeded case  approaches more to the QCRB than the coherent seeded case (Fig. \ref{fig_5}).

\section{Observation with Kerr state seeding as input in lossy case}\label{section4}

\begin{figure}
\includegraphics[width=8.5cm, height=5.5cm]{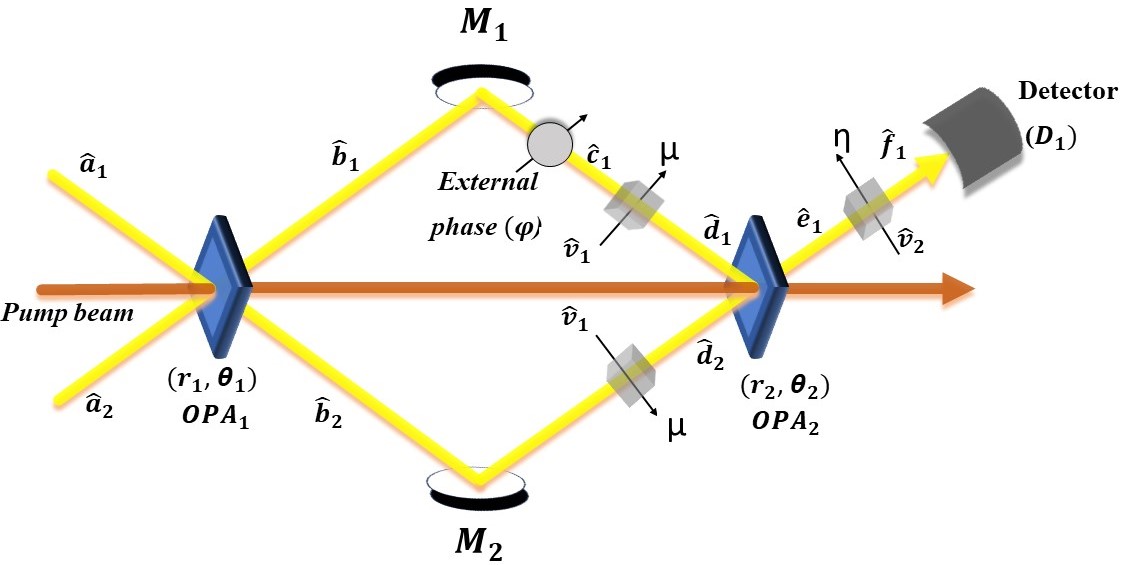}
\caption{\label{fig_7} Shows the schematic diagram of SU(1,1) interferometer in lossy case.}
 \end{figure}

In this section, we calculate the phase sensitivity in lossy cases using the Kerr seeding (Eq. \eqref{eqkerr}). In this case, the output operator $\hat{f}_1$ (Fig. \ref{fig_7}) can be written as
\begin{equation}
    \begin{split}
        \hat{f_1} = \hat{v}_2 \sqrt{1 - \eta} + \sqrt{\eta} \left(\sqrt{\mu} \cosh(r_2) \left(\hat{v}_1 \sqrt{\mu(1 - \mu)} \right. \right. \\
        + \left. e^{i\phi} \left(\hat{a}_1 e^{-2i\gamma \hat{a}_1^\dagger \hat{a}_1} \cosh(r_1) + e^{i\theta_1} \sinh(r_1) \hat{a}_2^\dagger \right)\right)\\
        + e^{i\theta_2} \sqrt{\mu} \sinh(r_2) \left(\hat{a}_1 e^{-i(\theta_1 + 2\gamma \hat{a}_1^\dagger \hat{a}_1)} \sinh(r_1) \right.\\
        \left. \left. + \cosh(r_1) \hat{a}_2^\dagger \right) + \sqrt{\mu(1 - \mu)} \hat{v}_1^\dagger \right)
    \end{split}
\end{equation}
Where, $\hat{a}_1$ is the annihilation operator for coherent state and $\hat{a}_2$, $\hat{v}_1$ and $\hat{v}_2$ are the annihilation operators for the vacuum state in different modes. Here, $\mu$ and $\eta$ are the internal and external loss coefficients respectivity ($\mu=\eta$=1, i.e, lossless case and $\mu = \eta\neq 1$, i.e., lossy case). 

In the lossy case, for the HD scheme, phase sensitivity reads as 
\begin{equation}
\Delta\phi'_{hd} = \frac{\sqrt{(\langle\hat{f}_1^{2}\rangle+\langle\hat{f}_1^{\dagger2}\rangle+2\langle\hat{f}_1^{\dagger}\hat{f}_1\rangle+1) -(\langle\hat{f}_1\rangle+\langle\hat{f}_1^\dagger\rangle)^2}}{\left|\frac{\partial(\langle\hat{f}_1\rangle+\langle\hat{f}_1^\dagger\rangle)}{\partial\phi}\right|}\label{da},
\end{equation}
where the values of $\langle\hat{f}_1\rangle,~\langle\hat{f}_1^{2}\rangle,~\langle\hat{f}_1^{\dagger}\hat{f}_1\rangle$ is given in Appendix \ref{appendix C}.

Since in the SI scheme, phase sensitivity did not depend on the $\gamma$, i.e. give the sensitivity equal to the coherent state, therefore discussion in the lossy case with the SI scheme will not give us any useful information. Therefore, here, we will discuss only the HD case.

\begin{figure}
\includegraphics[width=8.5cm, height=5.5cm]{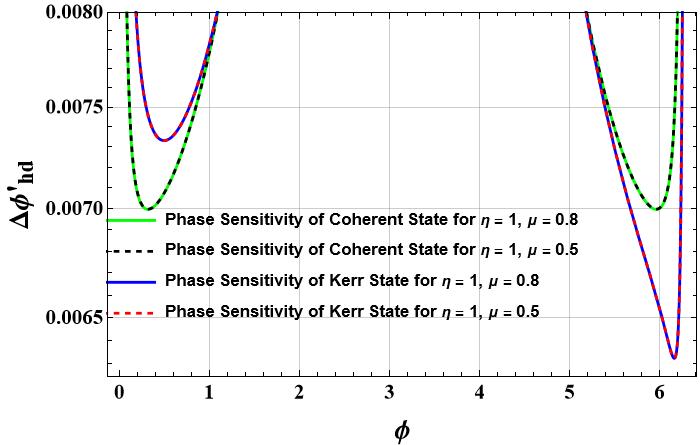}
\caption{\label{fig_9} The figure shows the variation of $\Delta\phi$ with $\phi$ for the internal loss with the values $r_1 = r_2 = 2.0, \gamma = 1\times10^{-6},~\theta_1=0,~\theta_2=\pi$, and $\alpha=100$. }
 \end{figure}

\begin{figure}
\includegraphics[width=8.5cm, height=5.5cm]{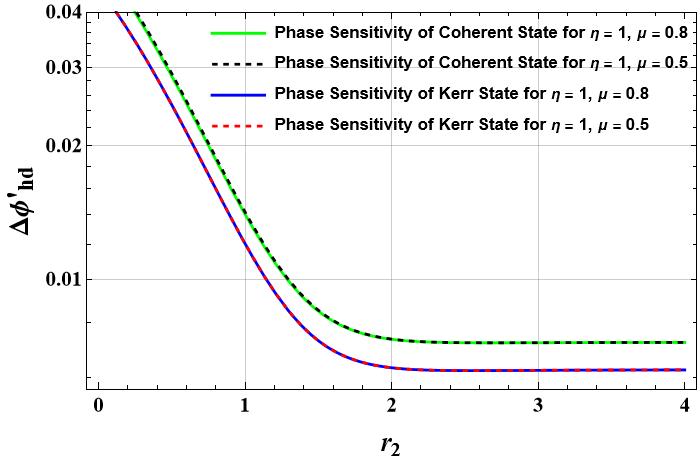}
\caption{\label{fig_10}Shows the variation of phase sensitivity $\Delta\phi$ versus $r_2$ for 
 the internal loss. The optimum values are $r_1 = 2.0, \gamma = 1\times10^{-6},~\theta_1=0,~\theta_2 = \pi,~\phi=6.15$, and $\alpha=100$.}
 \end{figure}

In the lossy case firstly we talk about the effect of internal loss ($\mu$) on phase sensitivity. Fig. \ref{fig_9} shows the effect of internal loss on the phase sensitivity. Interestingly, we can compensate the loss by improving the second OPA squeezing, $r_2$, (Fig. \ref{fig_10}).

\begin{figure}
\includegraphics[width=8.5cm, height=5.5cm]{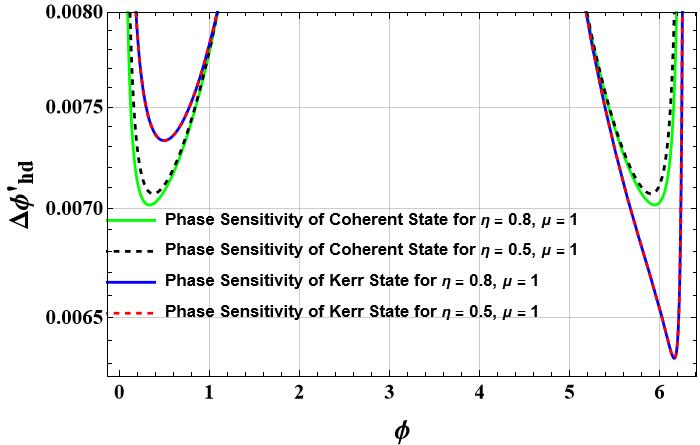}
\caption{\label{fig_11}The figure shows the variation of $\Delta\phi$ with $\phi$ for the external loss with the values $r_1 = r_2 = 2.0, \gamma = 1\times10^{-6},~\theta_1=0,~\theta_2=\pi$, and $\alpha=100$. }
 \end{figure}

\begin{figure}
\includegraphics[width=8.5cm, height=5.5cm]{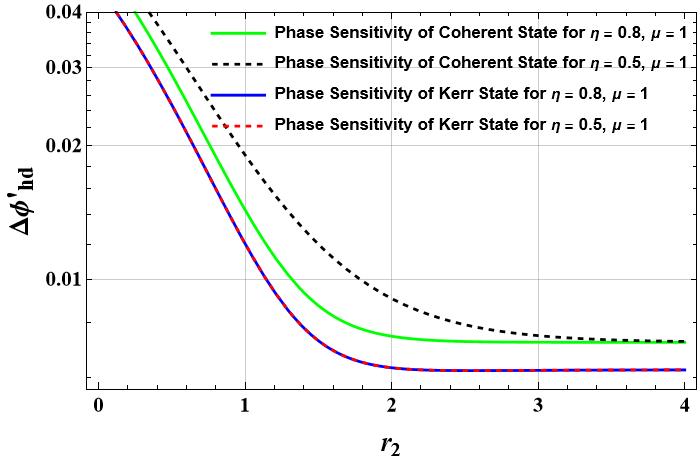}
\caption{\label{fig_12} Shows the variation of phase sensitivity $\Delta\phi$ versus $r_2$ for 
 the external loss. The optimum values are $r_1 = 2.0, \gamma = 1\times10^{-6},~\theta_1=0,~\theta_2 = \pi,~\phi=6.15$, and $\alpha=100$.}
 \end{figure}

To see the effect of external loss on the phase sensitivity we plot Fig. \ref{fig_11}. We find that by improving second OPA squeezing we can also compensate the external loss (Fig. \ref{fig_12}).

\section{Conclusions}\label{section5}

In this article, we have calculated the phase sensitivity of the SU(1,1) interferometer by using the Kerr state as seeding in lossless as well as in lossy conditions. For the detection scheme, we use SI and HD schemes. We have compared the results with the well-known coherent seeded case. 

In the HD scheme, we get the improvement in the phase sensitivity in the Kerr seeded case in comparison to the coherent seeded case. We found that the effect of loss (internal or external) can be compensated by improving the squeezing parameter of the second OPA. We also found that in the Kerr state case phase sensitivity is closer to the QCRB than the coherent state case. We have not get any improvement in the SI scheme with Kerr state seeding.

The physical reason behind the improvement of the sensitivity in the HD scheme basically depends on the squeezing possessed by the Kerr state. Since the squeezing ellipse in the Kerr state is oriented in an oblique direction (neither in the direction of phase quadrature nor the direction of amplitude quadrature) \cite{PhysRevA.34.3974}. After passing OPA squeezing increases in the Kerr state in comparison to a squeezed coherent state. Therefore, we get the advantage of inbuilt squeezing in the Kerr state in the phase sensitivity study in the HD scheme. However, the photon numbers are the same in both squeezed Kerr and squeezed coherent states therefore we did not get this improvement in the SI scheme.

In Summary, we can say that the Kerr state may be an alternative seeded state in SU(1,1) interferometer which is advantageous in quantum metrology and quantum sensing field.

\begin{acknowledgments}
PS, GS acknowledges UGC for the UGC Research Fellowship. GS acknowledges Incentive Grant under Institution of Eminence (IoE), Banaras Hindu University, Varanasi, India. DKM, AKP acknowledge financial support from the Science \& Engineering Research Board (SERB), New Delhi for CRG Grant (CRG/2021/005917). DKM acknowledges Incentive Grant under Institution of Eminence (IoE), Banaras Hindu University, Varanasi, India.
\end{acknowledgments}

\appendix

\section{Expectation values for lossless case}\label{appendix A}
\begin{widetext}
\begin{equation}
\langle\hat{d}_1\rangle = \alpha \left(1-2 i \alpha ^2 \gamma \right) (\sinh (r_1) \sinh (r_2) (\cos
   (\theta_1-\theta_2)-i \sin (\theta_1-\theta_2))+\cosh (r_1) \cosh
   (r_2) (\cos (\phi )+i \sin (\phi )))
\end{equation}

\begin{equation}
\langle\hat{d}_1^{2}\rangle =- \frac{1}{8} e^{-i (\theta_1 - \theta_2 - \phi)} \left(8 \alpha^2 (-1 + 2i \gamma + 4 \alpha^2 \gamma (i + 2\gamma + \alpha^2\gamma )) \left(\cos\left(\frac{\Phi}{2}\right) \cosh(r_1 + r_2) + i \cosh(r_1 - r_2) \sin\left(\frac{\Phi}{2}\right)\right)^2\right),
\end{equation}
\begin{equation}
\begin{split}
\langle\hat{d}_1^{\dagger}\hat{d}_1\rangle = \frac{1}{2} \left(-1 + \alpha^2  + 4 \alpha^4 (1 + \alpha^2) \gamma^2  + ((1 + \alpha^2) (1 + 4 \alpha^4 \gamma^2) \cosh(2r_1)) \cosh(r_2)^2 \right. \\
+ \left. ((1 + \alpha^2) (1 + 4 \alpha^4 \gamma^2) \cosh(2r_1)) \sinh(r_2)^2 + (1 + \alpha^2) (1 + 4 \alpha^4 \gamma^2) \cos(\phi) \sinh(2r_1) \sinh(2r_2)\right),
\end{split}
\end{equation}
where $\Phi = \theta_1 - \theta_2 + \phi$.
\end{widetext}
\section{Expectation values for QCRB}\label{appendix B}
\begin{widetext}
\begin{equation}
\begin{split}
\langle\Delta^2(\hat{c}_1^\dagger\hat{c}_1)\rangle =\frac{1}{2} \cosh(r1)^2 \left(-1 + 4 \alpha^4 \gamma^2 + 208 \alpha^8 \gamma^4 + 96 \alpha^{10} \gamma^4 + 8 \alpha^6 (\gamma^2 + 8 \gamma^4)\right.\\
+ \left.\left(1 + 2 \alpha^2 + 12 \alpha^4 \gamma^2 + 208 \alpha^8 \gamma^4 + 96 \alpha^{10} \gamma^4 + 16 \alpha^6 (\gamma^2 + 4 \gamma^4)\right) \cosh(2r_1)\right),
\end{split}
\end{equation}
\begin{equation}
\begin{split}
\langle\Delta^2(\hat{b}_2^\dagger\hat{b}_2)\rangle= \sinh(r_1)^2 \left(1 + 16 \alpha^4 \gamma^2 + 4 \alpha^6 \gamma^2 + \alpha^2 \left(1 + 8 \gamma^2\right)\right.+ \left(1 + 8 \gamma^2 + 1664 \alpha^8 \gamma^4 \right.\\
\left.\left.+ 192 \alpha^{10} \gamma^4 + 4 \alpha^4 \gamma^2 \left(37 + 752 \gamma^2\right) + 40 \alpha^6 \left(\gamma^2 + 104 \gamma^4\right) + \alpha^2 \left(2 + 96 \gamma^2 + 384 \gamma^4\right)\right) \sinh(r_1)^2\right),
\end{split}
\end{equation}
\begin{equation}
\begin{split}
Cov[(\hat{c}_1^\dagger\hat{c}_1),(\hat{b}_2^\dagger\hat{b}_2)]^2=\left(1 + 1280 \alpha^8 \gamma^4 + 192 \alpha^{10} \gamma^4 + 4 \alpha^4 \gamma^2 \left(25 + 272 \gamma^2\right) \right.\\\left.+ 8 \alpha^6 \gamma^2 \left(5 + 288 \gamma^2\right) + \alpha^2 \left(2 + 32 \gamma^2 + 64 \gamma^4\right)\right) \cosh(r_1)^2 \sinh(r_1)^2.
\end{split}
\end{equation}
\end{widetext}

\section{Expectation values for lossy case}\label{appendix C}
\begin{widetext}

\begin{equation}
    \langle\hat{f}_1\rangle = \alpha \sqrt{\eta } \sqrt{\mu } \left(1-2 i \alpha ^2 \gamma \right) (\sinh (r_1) \sinh (r_2) (\cos
   (\theta_1-\theta_2)-i \sin (\theta_1-\theta_2))+\cosh (r_1) \cosh
   (r_2) (\cos (\phi )+i \sin (\phi )))
\end{equation}

\begin{equation}
\begin{split}
\langle\hat{f}_1^{2}\rangle =- \frac{1}{8} \eta e^{-i (\theta_1 - \theta_2 - \phi)}  \left(8 \alpha^2 (-1 + 2i \gamma + 4 \alpha^2 \mu \gamma (i + (2 + \alpha^2) \gamma))  \left(\cos\left(\frac{\Phi}{2}\right) \cosh(r_1 + r_2) + i \cosh(r_1 - r_2) \sin\left(\frac{\Phi}{2}\right)\right)^2\right. \\
\left.+ 4 e^{i \theta_1} (-1 + \mu) \cos(\phi) \sinh(2r_2) + 4 (-1 + \mu) (-i \cos(\theta_1) + \sin(\theta_1)) \sin(\phi) \sinh(2r_2)\right),
\end{split}
\end{equation}
\begin{equation}
\begin{split}
\langle\hat{f}_1^{\dagger}\hat{f}_1\rangle = \frac{1}{2} \eta \left(-1 + \alpha^2 \mu + 4 \alpha^4 (1 + \alpha^2) \gamma^2 \mu + (1 - \mu + (1 + \alpha^2) (1 + 4 \alpha^4 \gamma^2) \mu \cosh(2r_1)) \cosh(r_2)^2 \right. \\
+ \left. (1 - \mu + (1 + \alpha^2) (1 + 4 \alpha^4 \gamma^2) \mu \cosh(2r_1)) \sinh(r_2)^2 + (1 + \alpha^2) (1 + 4 \alpha^4 \gamma^2) \mu \cos(\Phi) \sinh(2r_1) \sinh(2r_2)\right),
\end{split}
\end{equation}
where $\Phi = \theta_1 - \theta_2 + \phi$.
\end{widetext}
\bibliography{apssamp}

\begin{thebibliography}{28}%
\makeatletter
\providecommand \@ifxundefined [1]{%
 \@ifx{#1\undefined}
}%
\providecommand \@ifnum [1]{%
 \ifnum #1\expandafter \@firstoftwo
 \else \expandafter \@secondoftwo
 \fi
}%
\providecommand \@ifx [1]{%
 \ifx #1\expandafter \@firstoftwo
 \else \expandafter \@secondoftwo
 \fi
}%
\providecommand \natexlab [1]{#1}%
\providecommand \enquote  [1]{``#1''}%
\providecommand \bibnamefont  [1]{#1}%
\providecommand \bibfnamefont [1]{#1}%
\providecommand \citenamefont [1]{#1}%
\providecommand \href@noop [0]{\@secondoftwo}%
\providecommand \href [0]{\begingroup \@sanitize@url \@href}%
\providecommand \@href[1]{\@@startlink{#1}\@@href}%
\providecommand \@@href[1]{\endgroup#1\@@endlink}%
\providecommand \@sanitize@url [0]{\catcode `\\12\catcode `\$12\catcode `\&12\catcode `\#12\catcode `\^12\catcode `\_12\catcode `\%12\relax}%
\providecommand \@@startlink[1]{}%
\providecommand \@@endlink[0]{}%
\providecommand \url  [0]{\begingroup\@sanitize@url \@url }%
\providecommand \@url [1]{\endgroup\@href {#1}{\urlprefix }}%
\providecommand \urlprefix  [0]{URL }%
\providecommand \Eprint [0]{\href }%
\providecommand \doibase [0]{https://doi.org/}%
\providecommand \selectlanguage [0]{\@gobble}%
\providecommand \bibinfo  [0]{\@secondoftwo}%
\providecommand \bibfield  [0]{\@secondoftwo}%
\providecommand \translation [1]{[#1]}%
\providecommand \BibitemOpen [0]{}%
\providecommand \bibitemStop [0]{}%
\providecommand \bibitemNoStop [0]{.\EOS\space}%
\providecommand \EOS [0]{\spacefactor3000\relax}%
\providecommand \BibitemShut  [1]{\csname bibitem#1\endcsname}%
\let\auto@bib@innerbib\@empty
\bibitem [{\citenamefont {Helstrom}(1976)}]{1976Helstrom}%
  \BibitemOpen
  \bibfield  {author} {\bibinfo {author} {\bibfnamefont {C.~W.}\ \bibnamefont {Helstrom}},\ }\href {https://books.google.co.in/books?id=Ne3iT\_QLcsMC} {\emph {\bibinfo {title} {Quantum {D}etection and {E}stimation {T}heory}}}\ (\bibinfo  {publisher} {Academic Press, San Diego, CA},\ \bibinfo {year} {1976})\BibitemShut {NoStop}%
\bibitem [{\citenamefont {Dowling}(2008)}]{doi:10.1080/00107510802091298}%
  \BibitemOpen
  \bibfield  {author} {\bibinfo {author} {\bibfnamefont {J.~P.}\ \bibnamefont {Dowling}},\ }\bibfield  {title} {\bibinfo {title} {Quantum optical metrology – the lowdown on high-{N00N} states},\ }\href {https://doi.org/10.1080/00107510802091298} {\bibfield  {journal} {\bibinfo  {journal} {Contemporary Physics}\ }\textbf {\bibinfo {volume} {49}},\ \bibinfo {pages} {125} (\bibinfo {year} {2008})}\BibitemShut {NoStop}%
\bibitem [{\citenamefont {Pezz\`e}\ \emph {et~al.}(2018)\citenamefont {Pezz\`e}, \citenamefont {Smerzi}, \citenamefont {Oberthaler}, \citenamefont {Schmied},\ and\ \citenamefont {Treutlein}}]{RevModPhys.90.035005}%
  \BibitemOpen
  \bibfield  {author} {\bibinfo {author} {\bibfnamefont {L.}~\bibnamefont {Pezz\`e}}, \bibinfo {author} {\bibfnamefont {A.}~\bibnamefont {Smerzi}}, \bibinfo {author} {\bibfnamefont {M.~K.}\ \bibnamefont {Oberthaler}}, \bibinfo {author} {\bibfnamefont {R.}~\bibnamefont {Schmied}},\ and\ \bibinfo {author} {\bibfnamefont {P.}~\bibnamefont {Treutlein}},\ }\bibfield  {title} {\bibinfo {title} {Quantum metrology with nonclassical states of atomic ensembles},\ }\href {https://doi.org/10.1103/RevModPhys.90.035005} {\bibfield  {journal} {\bibinfo  {journal} {Rev. Mod. Phys.}\ }\textbf {\bibinfo {volume} {90}},\ \bibinfo {pages} {035005} (\bibinfo {year} {2018})}\BibitemShut {NoStop}%
\bibitem [{\citenamefont {{Demkowicz-Dobrzanski}}\ \emph {et~al.}(2015)\citenamefont {{Demkowicz-Dobrzanski}}, \citenamefont {{Jarzyna}},\ and\ \citenamefont {{Kolodynski}}}]{2015PrOpt..60..345D}%
  \BibitemOpen
  \bibfield  {author} {\bibinfo {author} {\bibfnamefont {R.}~\bibnamefont {{Demkowicz-Dobrzanski}}}, \bibinfo {author} {\bibfnamefont {M.}~\bibnamefont {{Jarzyna}}},\ and\ \bibinfo {author} {\bibfnamefont {J.}~\bibnamefont {{Kolodynski}}},\ }\bibfield  {title} {\bibinfo {title} {{Quantum limits in optical interferometry}},\ }\href {https://doi.org/10.1016/bs.po.2015.02.003} {\bibfield  {journal} {\bibinfo  {journal} {Progress in Optics}\ }\textbf {\bibinfo {volume} {60}},\ \bibinfo {pages} {345} (\bibinfo {year} {2015})},\ \Eprint {https://arxiv.org/abs/1405.7703} {arXiv:1405.7703 [quant-ph]} \BibitemShut {NoStop}%
\bibitem [{\citenamefont {Lawrie}\ \emph {et~al.}(2016)\citenamefont {Lawrie}, \citenamefont {Lett}, \citenamefont {Marino},\ and\ \citenamefont {Pooser}}]{Lawrie2019}%
  \BibitemOpen
  \bibfield  {author} {\bibinfo {author} {\bibfnamefont {B.~J.}\ \bibnamefont {Lawrie}}, \bibinfo {author} {\bibfnamefont {P.~D.}\ \bibnamefont {Lett}}, \bibinfo {author} {\bibfnamefont {A.~M.}\ \bibnamefont {Marino}},\ and\ \bibinfo {author} {\bibfnamefont {R.~C.}\ \bibnamefont {Pooser}},\ }\bibfield  {title} {\bibinfo {title} {Quantum {S}ensing with {S}queezed {L}ight},\ }\href {https://doi.org/10.1021/acsphotonics.9b00250} {\bibfield  {journal} {\bibinfo  {journal} {ACS Photonics}\ }\textbf {\bibinfo {volume} {6}},\ \bibinfo {pages} {1307} (\bibinfo {year} {2016})}\BibitemShut {NoStop}%
\bibitem [{\citenamefont {Sharma}\ \emph {et~al.}(2023)\citenamefont {Sharma}, \citenamefont {Mishra},\ and\ \citenamefont {Mishra}}]{sharma2023superresolution}%
  \BibitemOpen
  \bibfield  {author} {\bibinfo {author} {\bibfnamefont {P.}~\bibnamefont {Sharma}}, \bibinfo {author} {\bibfnamefont {M.~K.}\ \bibnamefont {Mishra}},\ and\ \bibinfo {author} {\bibfnamefont {D.~K.}\ \bibnamefont {Mishra}},\ }\href@noop {} {\bibinfo {title} {Super-resolution and super-sensitivity of quantum {LiDAR} with multi-photonic state and binary outcome photon counting measurement}} (\bibinfo {year} {2023}),\ \Eprint {https://arxiv.org/abs/2309.12076} {arXiv:2309.12076 [quant-ph]} \BibitemShut {NoStop}%
\bibitem [{\citenamefont {Shukla}\ \emph {et~al.}(2021)\citenamefont {Shukla}, \citenamefont {Salykina}, \citenamefont {Frascella}, \citenamefont {Mishra}, \citenamefont {Chekhova},\ and\ \citenamefont {Khalili}}]{Shukla:21}%
  \BibitemOpen
  \bibfield  {author} {\bibinfo {author} {\bibfnamefont {G.}~\bibnamefont {Shukla}}, \bibinfo {author} {\bibfnamefont {D.}~\bibnamefont {Salykina}}, \bibinfo {author} {\bibfnamefont {G.}~\bibnamefont {Frascella}}, \bibinfo {author} {\bibfnamefont {D.~K.}\ \bibnamefont {Mishra}}, \bibinfo {author} {\bibfnamefont {M.~V.}\ \bibnamefont {Chekhova}},\ and\ \bibinfo {author} {\bibfnamefont {F.~Y.}\ \bibnamefont {Khalili}},\ }\bibfield  {title} {\bibinfo {title} {Broadening the high sensitivity range of squeezing-assisted interferometers by means of two-channel detection},\ }\href {https://doi.org/10.1364/OE.413391} {\bibfield  {journal} {\bibinfo  {journal} {Opt. Express}\ }\textbf {\bibinfo {volume} {29}},\ \bibinfo {pages} {95} (\bibinfo {year} {2021})}\BibitemShut {NoStop}%
\bibitem [{\citenamefont {Shukla}\ \emph {et~al.}(2022)\citenamefont {Shukla}, \citenamefont {Mishra}, \citenamefont {Yadav}, \citenamefont {Pandey},\ and\ \citenamefont {Mishra}}]{Shukla:22}%
  \BibitemOpen
  \bibfield  {author} {\bibinfo {author} {\bibfnamefont {G.}~\bibnamefont {Shukla}}, \bibinfo {author} {\bibfnamefont {K.~K.}\ \bibnamefont {Mishra}}, \bibinfo {author} {\bibfnamefont {D.}~\bibnamefont {Yadav}}, \bibinfo {author} {\bibfnamefont {R.~K.}\ \bibnamefont {Pandey}},\ and\ \bibinfo {author} {\bibfnamefont {D.~K.}\ \bibnamefont {Mishra}},\ }\bibfield  {title} {\bibinfo {title} {Quantum-enhanced super-sensitivity of a {M}ach--{Z}ehnder interferometer with superposition of {S}chr\"{o}dinger's cat-like state and {F}ock state as inputs using a two-channel detection},\ }\href {https://doi.org/10.1364/JOSAB.434967} {\bibfield  {journal} {\bibinfo  {journal} {J. Opt. Soc. Am. B}\ }\textbf {\bibinfo {volume} {39}},\ \bibinfo {pages} {59} (\bibinfo {year} {2022})}\BibitemShut {NoStop}%
\bibitem [{\citenamefont {Mishra}\ \emph {et~al.}(2021)\citenamefont {Mishra}, \citenamefont {Yadav}, \citenamefont {Shukla},\ and\ \citenamefont {Mishra}}]{Mishra_2021}%
  \BibitemOpen
  \bibfield  {author} {\bibinfo {author} {\bibfnamefont {K.~K.}\ \bibnamefont {Mishra}}, \bibinfo {author} {\bibfnamefont {D.}~\bibnamefont {Yadav}}, \bibinfo {author} {\bibfnamefont {G.}~\bibnamefont {Shukla}},\ and\ \bibinfo {author} {\bibfnamefont {D.~K.}\ \bibnamefont {Mishra}},\ }\bibfield  {title} {\bibinfo {title} {Non-classicalities exhibited by the superposition of {S}chrödinger's cat state with the vacuum of the optical field},\ }\href {https://doi.org/10.1088/1402-4896/abe00f} {\bibfield  {journal} {\bibinfo  {journal} {Physica Scripta}\ }\textbf {\bibinfo {volume} {96}},\ \bibinfo {pages} {045102} (\bibinfo {year} {2021})}\BibitemShut {NoStop}%
\bibitem [{\citenamefont {Shukla}\ \emph {et~al.}(2024)\citenamefont {Shukla}, \citenamefont {Yadav}, \citenamefont {Sharma}, \citenamefont {Kumar},\ and\ \citenamefont {Mishra}}]{SHUKLA2024100200}%
  \BibitemOpen
  \bibfield  {author} {\bibinfo {author} {\bibfnamefont {G.}~\bibnamefont {Shukla}}, \bibinfo {author} {\bibfnamefont {D.}~\bibnamefont {Yadav}}, \bibinfo {author} {\bibfnamefont {P.}~\bibnamefont {Sharma}}, \bibinfo {author} {\bibfnamefont {A.}~\bibnamefont {Kumar}},\ and\ \bibinfo {author} {\bibfnamefont {D.~K.}\ \bibnamefont {Mishra}},\ }\bibfield  {title} {\bibinfo {title} {Quantum sub-phase sensitivity of a mach–zehnder interferometer with the superposition of schrödinger’s cat-like state with vacuum state as an input under product detection scheme},\ }\href {https://doi.org/https://doi.org/10.1016/j.physo.2023.100200} {\bibfield  {journal} {\bibinfo  {journal} {Physics Open}\ }\textbf {\bibinfo {volume} {18}},\ \bibinfo {pages} {100200} (\bibinfo {year} {2024})}\BibitemShut {NoStop}%
\bibitem [{\citenamefont {Shukla}\ \emph {et~al.}(2023)\citenamefont {Shukla}, \citenamefont {Mishra}, \citenamefont {Pandey}, \citenamefont {Kumar}, \citenamefont {Pandey},\ and\ \citenamefont {Mishra}}]{shukla2023improvement}%
  \BibitemOpen
  \bibfield  {author} {\bibinfo {author} {\bibfnamefont {G.}~\bibnamefont {Shukla}}, \bibinfo {author} {\bibfnamefont {K.~M.}\ \bibnamefont {Mishra}}, \bibinfo {author} {\bibfnamefont {A.~K.}\ \bibnamefont {Pandey}}, \bibinfo {author} {\bibfnamefont {T.}~\bibnamefont {Kumar}}, \bibinfo {author} {\bibfnamefont {H.}~\bibnamefont {Pandey}},\ and\ \bibinfo {author} {\bibfnamefont {D.~K.}\ \bibnamefont {Mishra}},\ }\bibfield  {title} {\bibinfo {title} {Improvement in phase-sensitivity of a mach--zehnder interferometer with the superposition of schr{\"o}dinger’s cat-like state with vacuum state as an input under parity measurement},\ }\href {https://doi.org/10.1007/s11082-023-04724-w} {\bibfield  {journal} {\bibinfo  {journal} {Optical and Quantum Electronics}\ }\textbf {\bibinfo {volume} {55}},\ \bibinfo {pages} {460} (\bibinfo {year} {2023})}\BibitemShut {NoStop}%
\bibitem [{\citenamefont {Chekhova}\ and\ \citenamefont {Ou}(2016)}]{Chekhova:16}%
  \BibitemOpen
  \bibfield  {author} {\bibinfo {author} {\bibfnamefont {M.~V.}\ \bibnamefont {Chekhova}}\ and\ \bibinfo {author} {\bibfnamefont {Z.~Y.}\ \bibnamefont {Ou}},\ }\bibfield  {title} {\bibinfo {title} {Nonlinear interferometers in quantum optics},\ }\href {https://doi.org/10.1364/AOP.8.000104} {\bibfield  {journal} {\bibinfo  {journal} {Adv. Opt. Photon.}\ }\textbf {\bibinfo {volume} {8}},\ \bibinfo {pages} {104} (\bibinfo {year} {2016})}\BibitemShut {NoStop}%
\bibitem [{\citenamefont {Li}\ \emph {et~al.}(2016)\citenamefont {Li}, \citenamefont {Gard}, \citenamefont {Gao}, \citenamefont {Yuan}, \citenamefont {Zhang}, \citenamefont {Lee},\ and\ \citenamefont {Dowling}}]{PhysRevA.94.063840}%
  \BibitemOpen
  \bibfield  {author} {\bibinfo {author} {\bibfnamefont {D.}~\bibnamefont {Li}}, \bibinfo {author} {\bibfnamefont {B.~T.}\ \bibnamefont {Gard}}, \bibinfo {author} {\bibfnamefont {Y.}~\bibnamefont {Gao}}, \bibinfo {author} {\bibfnamefont {C.-H.}\ \bibnamefont {Yuan}}, \bibinfo {author} {\bibfnamefont {W.}~\bibnamefont {Zhang}}, \bibinfo {author} {\bibfnamefont {H.}~\bibnamefont {Lee}},\ and\ \bibinfo {author} {\bibfnamefont {J.~P.}\ \bibnamefont {Dowling}},\ }\bibfield  {title} {\bibinfo {title} {Phase sensitivity at the {H}eisenberg limit in an {SU}(1,1) interferometer via parity detection},\ }\href {https://doi.org/10.1103/PhysRevA.94.063840} {\bibfield  {journal} {\bibinfo  {journal} {Phys. Rev. A}\ }\textbf {\bibinfo {volume} {94}},\ \bibinfo {pages} {063840} (\bibinfo {year} {2016})}\BibitemShut {NoStop}%
\bibitem [{\citenamefont {Yurke}\ \emph {et~al.}(1986)\citenamefont {Yurke}, \citenamefont {McCall},\ and\ \citenamefont {Klauder}}]{PhysRevA.33.4033}%
  \BibitemOpen
  \bibfield  {author} {\bibinfo {author} {\bibfnamefont {B.}~\bibnamefont {Yurke}}, \bibinfo {author} {\bibfnamefont {S.~L.}\ \bibnamefont {McCall}},\ and\ \bibinfo {author} {\bibfnamefont {J.~R.}\ \bibnamefont {Klauder}},\ }\bibfield  {title} {\bibinfo {title} {{S}{U}(2) and {S}{U}(1,1) interferometers},\ }\href {https://doi.org/10.1103/PhysRevA.33.4033} {\bibfield  {journal} {\bibinfo  {journal} {Phys. Rev. A}\ }\textbf {\bibinfo {volume} {33}},\ \bibinfo {pages} {4033} (\bibinfo {year} {1986})}\BibitemShut {NoStop}%
\bibitem [{\citenamefont {Plick}\ \emph {et~al.}(2010)\citenamefont {Plick}, \citenamefont {Dowling},\ and\ \citenamefont {Agarwal}}]{Plick1_2010}%
  \BibitemOpen
  \bibfield  {author} {\bibinfo {author} {\bibfnamefont {W.~N.}\ \bibnamefont {Plick}}, \bibinfo {author} {\bibfnamefont {J.~P.}\ \bibnamefont {Dowling}},\ and\ \bibinfo {author} {\bibfnamefont {G.~S.}\ \bibnamefont {Agarwal}},\ }\bibfield  {title} {\bibinfo {title} {Coherent-light-boosted, sub-shot noise, quantum interferometry},\ }\href {https://doi.org/10.1088/1367-2630/12/8/083014} {\bibfield  {journal} {\bibinfo  {journal} {New Journal of Physics}\ }\textbf {\bibinfo {volume} {12}},\ \bibinfo {pages} {083014} (\bibinfo {year} {2010})}\BibitemShut {NoStop}%
\bibitem [{\citenamefont {Marino}\ \emph {et~al.}(2012)\citenamefont {Marino}, \citenamefont {Corzo~Trejo},\ and\ \citenamefont {Lett}}]{PhysRevA.86.023844}%
  \BibitemOpen
  \bibfield  {author} {\bibinfo {author} {\bibfnamefont {A.~M.}\ \bibnamefont {Marino}}, \bibinfo {author} {\bibfnamefont {N.~V.}\ \bibnamefont {Corzo~Trejo}},\ and\ \bibinfo {author} {\bibfnamefont {P.~D.}\ \bibnamefont {Lett}},\ }\bibfield  {title} {\bibinfo {title} {Effect of losses on the performance of an {SU}(1,1) interferometer},\ }\href {https://doi.org/10.1103/PhysRevA.86.023844} {\bibfield  {journal} {\bibinfo  {journal} {Phys. Rev. A}\ }\textbf {\bibinfo {volume} {86}},\ \bibinfo {pages} {023844} (\bibinfo {year} {2012})}\BibitemShut {NoStop}%
\bibitem [{\citenamefont {Hudelist}\ \emph {et~al.}(2014)\citenamefont {Hudelist}, \citenamefont {Kong}, \citenamefont {Liu}, \citenamefont {Jing}, \citenamefont {Ou},\ and\ \citenamefont {Zhang}}]{hudelist2014quantum}%
  \BibitemOpen
  \bibfield  {author} {\bibinfo {author} {\bibfnamefont {F.}~\bibnamefont {Hudelist}}, \bibinfo {author} {\bibfnamefont {J.}~\bibnamefont {Kong}}, \bibinfo {author} {\bibfnamefont {C.}~\bibnamefont {Liu}}, \bibinfo {author} {\bibfnamefont {J.}~\bibnamefont {Jing}}, \bibinfo {author} {\bibfnamefont {Z.}~\bibnamefont {Ou}},\ and\ \bibinfo {author} {\bibfnamefont {W.}~\bibnamefont {Zhang}},\ }\bibfield  {title} {\bibinfo {title} {Quantum metrology with parametric amplifier-based photon correlation interferometers},\ }\href {https://doi.org/10.1038/ncomms4049} {\bibfield  {journal} {\bibinfo  {journal} {Nature communications}\ }\textbf {\bibinfo {volume} {5}},\ \bibinfo {pages} {3049} (\bibinfo {year} {2014})}\BibitemShut {NoStop}%
\bibitem [{\citenamefont {Chang}\ \emph {et~al.}(2022)\citenamefont {Chang}, \citenamefont {Ye}, \citenamefont {Zhang}, \citenamefont {Hu}, \citenamefont {Huang},\ and\ \citenamefont {Liu}}]{PhysRevA.105.033704}%
  \BibitemOpen
  \bibfield  {author} {\bibinfo {author} {\bibfnamefont {S.}~\bibnamefont {Chang}}, \bibinfo {author} {\bibfnamefont {W.}~\bibnamefont {Ye}}, \bibinfo {author} {\bibfnamefont {H.}~\bibnamefont {Zhang}}, \bibinfo {author} {\bibfnamefont {L.}~\bibnamefont {Hu}}, \bibinfo {author} {\bibfnamefont {J.}~\bibnamefont {Huang}},\ and\ \bibinfo {author} {\bibfnamefont {S.}~\bibnamefont {Liu}},\ }\bibfield  {title} {\bibinfo {title} {Improvement of phase sensitivity in an {SU}(1,1) interferometer via a phase shift induced by a {K}err medium},\ }\href {https://doi.org/10.1103/PhysRevA.105.033704} {\bibfield  {journal} {\bibinfo  {journal} {Phys. Rev. A}\ }\textbf {\bibinfo {volume} {105}},\ \bibinfo {pages} {033704} (\bibinfo {year} {2022})}\BibitemShut {NoStop}%
\bibitem [{\citenamefont {Yadav}\ \emph {et~al.}(2023)\citenamefont {Yadav}, \citenamefont {Shukla}, \citenamefont {Sharma},\ and\ \citenamefont {Mishra}}]{yadav2023quantumenhanced}%
  \BibitemOpen
  \bibfield  {author} {\bibinfo {author} {\bibfnamefont {D.}~\bibnamefont {Yadav}}, \bibinfo {author} {\bibfnamefont {G.}~\bibnamefont {Shukla}}, \bibinfo {author} {\bibfnamefont {P.}~\bibnamefont {Sharma}},\ and\ \bibinfo {author} {\bibfnamefont {D.~K.}\ \bibnamefont {Mishra}},\ }\href@noop {} {\bibinfo {title} {Quantum-enhanced super-sensitivity of {M}ach-{Z}ehnder interferometer using squeezed {K}err state}} (\bibinfo {year} {2023}),\ \Eprint {https://arxiv.org/abs/2309.04731} {arXiv:2309.04731 [quant-ph]} \BibitemShut {NoStop}%
\bibitem [{\citenamefont {Kitagawa}\ and\ \citenamefont {Yamamoto}(1986)}]{PhysRevA.34.3974}%
  \BibitemOpen
  \bibfield  {author} {\bibinfo {author} {\bibfnamefont {M.}~\bibnamefont {Kitagawa}}\ and\ \bibinfo {author} {\bibfnamefont {Y.}~\bibnamefont {Yamamoto}},\ }\bibfield  {title} {\bibinfo {title} {Number-phase minimum-uncertainty state with reduced number uncertainty in a {K}err nonlinear interferometer},\ }\href {https://doi.org/10.1103/PhysRevA.34.3974} {\bibfield  {journal} {\bibinfo  {journal} {Phys. Rev. A}\ }\textbf {\bibinfo {volume} {34}},\ \bibinfo {pages} {3974} (\bibinfo {year} {1986})}\BibitemShut {NoStop}%
\bibitem [{\citenamefont {Sizmann}\ and\ \citenamefont {Leuchs}(1999)}]{sizmann1999v}%
  \BibitemOpen
  \bibfield  {author} {\bibinfo {author} {\bibfnamefont {A.}~\bibnamefont {Sizmann}}\ and\ \bibinfo {author} {\bibfnamefont {G.}~\bibnamefont {Leuchs}},\ }\bibfield  {title} {\bibinfo {title} {V the optical {K}err effect and quantum optics in fibers},\ }\href@noop {} {\bibfield  {journal} {\bibinfo  {journal} {Progress in optics}\ }\textbf {\bibinfo {volume} {39}},\ \bibinfo {pages} {373} (\bibinfo {year} {1999})}\BibitemShut {NoStop}%
\bibitem [{\citenamefont {You}\ \emph {et~al.}(2019)\citenamefont {You}, \citenamefont {Adhikari}, \citenamefont {Ma}, \citenamefont {Sasaki}, \citenamefont {Takeoka},\ and\ \citenamefont {Dowling}}]{PhysRevA.99.042122}%
  \BibitemOpen
  \bibfield  {author} {\bibinfo {author} {\bibfnamefont {C.}~\bibnamefont {You}}, \bibinfo {author} {\bibfnamefont {S.}~\bibnamefont {Adhikari}}, \bibinfo {author} {\bibfnamefont {X.}~\bibnamefont {Ma}}, \bibinfo {author} {\bibfnamefont {M.}~\bibnamefont {Sasaki}}, \bibinfo {author} {\bibfnamefont {M.}~\bibnamefont {Takeoka}},\ and\ \bibinfo {author} {\bibfnamefont {J.~P.}\ \bibnamefont {Dowling}},\ }\bibfield  {title} {\bibinfo {title} {Conclusive precision bounds for su(1,1) interferometers},\ }\href {https://doi.org/10.1103/PhysRevA.99.042122} {\bibfield  {journal} {\bibinfo  {journal} {Phys. Rev. A}\ }\textbf {\bibinfo {volume} {99}},\ \bibinfo {pages} {042122} (\bibinfo {year} {2019})}\BibitemShut {NoStop}%
\bibitem [{\citenamefont {Zeng}\ \emph {et~al.}(2023)\citenamefont {Zeng}, \citenamefont {Ding}, \citenamefont {Zhou}, \citenamefont {Jiao}, \citenamefont {Zhang}, \citenamefont {Chen}, \citenamefont {Zhang},\ and\ \citenamefont {Yuan}}]{zeng2023effect}%
  \BibitemOpen
  \bibfield  {author} {\bibinfo {author} {\bibfnamefont {J.}~\bibnamefont {Zeng}}, \bibinfo {author} {\bibfnamefont {Y.}~\bibnamefont {Ding}}, \bibinfo {author} {\bibfnamefont {M.}~\bibnamefont {Zhou}}, \bibinfo {author} {\bibfnamefont {G.-F.}\ \bibnamefont {Jiao}}, \bibinfo {author} {\bibfnamefont {K.}~\bibnamefont {Zhang}}, \bibinfo {author} {\bibfnamefont {L.~Q.}\ \bibnamefont {Chen}}, \bibinfo {author} {\bibfnamefont {W.}~\bibnamefont {Zhang}},\ and\ \bibinfo {author} {\bibfnamefont {C.-H.}\ \bibnamefont {Yuan}},\ }\href@noop {} {\bibinfo {title} {The effect of {Q}uantum {S}tatistics on the sensitivity in an {SU}(1,1) interferometer}} (\bibinfo {year} {2023}),\ \Eprint {https://arxiv.org/abs/2308.03002} {arXiv:2308.03002 [quant-ph]} \BibitemShut {NoStop}%
\bibitem [{\citenamefont {Agarwal}(2012)}]{Agarwal_2012}%
  \BibitemOpen
  \bibfield  {author} {\bibinfo {author} {\bibfnamefont {G.~S.}\ \bibnamefont {Agarwal}},\ }\href {https://doi.org/10.1017/cbo9781139035170} {\emph {\bibinfo {title} {Quantum Optics}}}\ (\bibinfo  {publisher} {Cambridge University Press},\ \bibinfo {address} {Cambridge, England},\ \bibinfo {year} {2012})\BibitemShut {NoStop}%
\bibitem [{\citenamefont {Gerry}\ and\ \citenamefont {Knight}(2004)}]{gerry_knight_2004}%
  \BibitemOpen
  \bibfield  {author} {\bibinfo {author} {\bibfnamefont {C.~C.}\ \bibnamefont {Gerry}}\ and\ \bibinfo {author} {\bibfnamefont {P.~L.}\ \bibnamefont {Knight}},\ }\href {https://doi.org/10.1017/CBO9780511791239} {\emph {\bibinfo {title} {Introductory Quantum Optics}}}\ (\bibinfo  {publisher} {Cambridge University Press},\ \bibinfo {year} {2004})\BibitemShut {NoStop}%
\bibitem [{\citenamefont {Bunker}\ \emph {et~al.}(2019)\citenamefont {Bunker}, \citenamefont {Mills},\ and\ \citenamefont {Jensen}}]{BUNKER2019106594}%
  \BibitemOpen
  \bibfield  {author} {\bibinfo {author} {\bibfnamefont {P.}~\bibnamefont {Bunker}}, \bibinfo {author} {\bibfnamefont {I.~M.}\ \bibnamefont {Mills}},\ and\ \bibinfo {author} {\bibfnamefont {P.}~\bibnamefont {Jensen}},\ }\bibfield  {title} {\bibinfo {title} {The {P}lanck constant and its units},\ }\href {https://doi.org/https://doi.org/10.1016/j.jqsrt.2019.106594} {\bibfield  {journal} {\bibinfo  {journal} {Journal of Quantitative Spectroscopy and Radiative Transfer}\ }\textbf {\bibinfo {volume} {237}},\ \bibinfo {pages} {106594} (\bibinfo {year} {2019})}\BibitemShut {NoStop}%
\bibitem [{\citenamefont {Lang}\ and\ \citenamefont {Caves}(2013)}]{PhysRevLett.111.173601}%
  \BibitemOpen
  \bibfield  {author} {\bibinfo {author} {\bibfnamefont {M.~D.}\ \bibnamefont {Lang}}\ and\ \bibinfo {author} {\bibfnamefont {C.~M.}\ \bibnamefont {Caves}},\ }\bibfield  {title} {\bibinfo {title} {Optimal {Q}uantum-{E}nhanced {I}nterferometry {U}sing a {L}aser {P}ower {S}ource},\ }\href {https://doi.org/10.1103/PhysRevLett.111.173601} {\bibfield  {journal} {\bibinfo  {journal} {Phys. Rev. Lett.}\ }\textbf {\bibinfo {volume} {111}},\ \bibinfo {pages} {173601} (\bibinfo {year} {2013})}\BibitemShut {NoStop}%
\bibitem [{\citenamefont {Gerry}\ and\ \citenamefont {Grobe}(1994)}]{PhysRevA.49.2033}%
  \BibitemOpen
  \bibfield  {author} {\bibinfo {author} {\bibfnamefont {C.~C.}\ \bibnamefont {Gerry}}\ and\ \bibinfo {author} {\bibfnamefont {R.}~\bibnamefont {Grobe}},\ }\bibfield  {title} {\bibinfo {title} {Statistical properties of squeezed kerr states},\ }\href {https://doi.org/10.1103/PhysRevA.49.2033} {\bibfield  {journal} {\bibinfo  {journal} {Phys. Rev. A}\ }\textbf {\bibinfo {volume} {49}},\ \bibinfo {pages} {2033} (\bibinfo {year} {1994})}\BibitemShut {NoStop}%
\end{thebibliography}%

\end{document}